\documentclass{iosart2c}
\usepackage[T1]{fontenc}
\usepackage{times}%
\usepackage{amsmath}
\usepackage{dcolumn}
\usepackage{graphics}
\usepackage{amsmath,amsfonts,amssymb,epsfig,epstopdf,url,array}

\usepackage{amsthm}
\usepackage{color,soul}
\usepackage{algorithmic}
\usepackage{balance}
\usepackage{stmaryrd}
\usepackage[table]{xcolor}
\usepackage{centernot}
\usepackage{tikz}
\usepackage{booktabs}
\usepackage{pdfpages}
\usepackage{graphicx}
\usepackage{subfigure}
\usepackage[font=small,skip=8pt]{caption}
\usepackage{tabularx}
\usepackage{comment}
\usepackage{multirow}
\usepackage{multicol}
\usepackage{bigstrut}
\usepackage{pdflscape}
\usepackage{rotating}
\usepackage{float}
\usepackage{lipsum}
\usepackage{enumitem}
\usetikzlibrary{plotmarks,shapes,arrows,chains,hobby,backgrounds,calc,trees}
\usepackage{makecell}
\usepackage{dblfloatfix}    
\usepackage{pdfcomment} 
\usepackage{arydshln}
\usepackage{amssymb}
\usepackage{bm}
\usepackage{arydshln}
\usepackage{lipsum}                     
\usepackage{xargs}                      

\usepackage[colorinlistoftodos,prependcaption,textsize=tiny]{todonotes}

\newcommandx{\unsure}[2][1=]{\todo[linecolor=red,backgroundcolor=red!25,bordercolor=red,#1]{#2}}
\newcommandx{\change}[2][1=]{\todo[linecolor=blue,backgroundcolor=blue!25,bordercolor=blue,#1]{#2}}
\newcommandx{\info}[2][1=]{\todo[linecolor=OliveGreen,backgroundcolor=OliveGreen!25,bordercolor=OliveGreen,#1]{#2}}
\newcommandx{\improvement}[2][1=]{\todo[linecolor=orange,backgroundcolor=orange!25,bordercolor=orange,#1]{#2}}
\newcommandx{\hiddencomment}[2][1=]{\todo[disable,#1]{#2}}
\newcommand*{\Perm}[2]{{}^{#1}\!P_{#2}}%
\newcolumntype{d}[1]{D{.}{.}{#1}}

\definecolor{lightgray}{gray}{0.9}
\theoremstyle{definition}
\newtheorem{exmp}{Example}
\theoremstyle{definition}

\theoremstyle{definition}
\newtheorem{defn}{Definition}

\firstpage{1} \lastpage{15} \volume{1} \pubyear{2020}
\begin{document}
\begin{frontmatter}                           

%
\title{Dependency-Aware Release Planning\\ for Software Projects  using \\Fuzzy Graphs and Integer Programming}

\runningtitle{Dependency-Aware Software Release Planning using Fuzzy Graphs and Integer Programming}

\author[A]{\fnms{Davoud} \snm{Mougouei}\thanks{Corresponding author. E-mail: dmougouei@gmail.com}} and 
\author[B]{\fnms{David} \snm{M W Powers}}
\runningauthor{D. Mougouei et al.}
\address[A]{Faculty of IT,  Monash University, VIC,
	Australia \\E-mail: Davoud.Mougouei@Monash.edu}

\address[B]{College of Science and Engineering,  Flinders University, SA,
Australia\\ E-mail: David.Powers@Flinders.edu.au}

\vspace{-0.8cm}
\begin{abstract}
\textit{Software Release Planning} (SRP) is to find, for the software, a subset of the requirements with the highest value while respecting the budget. The value of a requirement however may, to various degrees, depend on selecting or ignoring other requirements. However, existing SRP models ignore either \textit{Value-Related Dependencies} altogether or the strengths of those dependencies even if they consider them. This paper presents an \textit{Integer Programming} model for software release planning, which considers the variances of strengths of positive and negative value-related dependencies among software requirements. To capture the imprecision associated with strengths of value-related dependencies we have made use of the algebraic structure of fuzzy graphs. We have further, contributed a scalable technique for automated identification of value-related dependencies based on user preferences for software requirements. The validity of the work is verified through simulations.

\end{abstract}

\begin{keyword}
Fuzzy Graphs\sep Integer Programming\sep Value-Related Dependencies\sep Release Planning
\end{keyword}

\end{frontmatter}

\section{Introduction}
\label{sec_introduction}

Owing to budget constraints, software release planning (SRP)~\cite{franch2016software,bagnall_next_2001} is needed to find a subset of software requirements (optimal set) with the highest value while keeping the cost of the selected requirements within the available budget~\cite{dahlstedt_moulding_2003}. The value of a requirement however, may be positively or negatively influenced by selecting or ignoring other requirements due to the dependencies among values of requirements~\cite{mougouei2020dependency,mougouei2018phd,mougoueifuzzy,Zhang_RIM_2013,Robinson_RIM_2003,dahlstedt2005requirements,mougouei2019dependency}. Value-related dependencies~\cite{carlshamre_industrial_2001,li_integrated_2010}, also known as \textit{Increases-Decreases-value-of}~\cite{dahlstedt2005requirements,zhang_investigating_2014}, \textit{CVALUE}~\cite{carlshamre_industrial_2001}, and \textit{Positive-Negative value dependencies}~\cite{karlsson_improved_1997} are commonly found in software projects~\cite{carlshamre_industrial_2001,carlshamre_release_2002,pitangueira2015software}. Requirement dependencies, including value-related requirement dependencies, on the other hand, are fuzzy relations~\cite{carlshamre_industrial_2001,ngo_fuzzy_2005,tang_using_2007} meaning that the strengths of those dependencies are imprecise and vary from large to insignificant in the context of real software projects~\cite{Robinson_RIM_2003,ngo2005measuring,carlshamre_industrial_2001,wang_simulation_2012}. It is important thus, to consider not only the existence but also the strengths of value-related dependencies~\cite{dahlstedt2005requirements, brasil_multiobjective_2012,harman_exact_2014} in software release planning. Carlshamre \textit{et al.}~\cite{carlshamre_industrial_2001} observed the need to consider the strengths of requirement dependencies but they did not go further on how to formally achieve this. However, the existing software release planning models either ignore requirement dependencies~\cite{karlsson_optimizing_1997,jung_optimizing_1998,ruhe_trade_2003} or do not consider the variance of strengths of those dependencies~\cite{li_integrated_2010,brasil_multiobjective_2012,bagnall_next_2001,baker_search_2006} by treating them as precede relations~\cite{veerapen2015integer,Martello_BKP_1990}. 
Following the latter direction, ignoring (selecting) a requirement $r_i$ results in ignoring all of the requirements whose values positively (negatively) depend on $r_i$ even if budget allows for their implementation~\cite{li_integrated_2010}. 

This results in a  condition referred to as the \textit{Selection Deficiency Problem} (SDP)~\cite{mougouei2016factoring}. As a result of the SDP, any increase in the number of value-related dependencies will significantly reduce the value of the selected requirements~\cite{li_integrated_2010}. 

To avoid these issues and properly factor value-related dependencies in software release planning, We proposed the notion of ``Dependency-Aware Software Release Planning'' in our earlier works~\cite{mougouei2017dasrp,mougouei2017preference} to account for value-related dependencies. To achieve this, we have presented four major contributions as follows. 

First, we have formulated an integer programming~\cite{wolsey1998integer} model referred to as the \textit{Dependency-Aware Software Release Planning} (DA-SRP) that finds a subset of software requirements with the highest \textit{Overall Value} (OV), where overall value captures the influences of value-related dependencies (derived from user preferences) on the value of a selected subset of requirements. The DA-SRP model considers qualities (positivity or negativity) and strengths of value-related requirement dependencies.

Second, we have introduced \textit{Value Dependency Graphs} (VDGs) based on fuzzy graphs~\cite{kalampakas_fuzzy_2013,rosenfeld_fuzzygraph_1975} for modeling value-related requirement dependencies and capturing the imprecision associated with strengths of those dependencies. A modified version of the Floyd-Warshall algorithm~\cite{floyd_1962} is devised to infer implicit value-related dependencies in polynomial time. 

Third, we have contributed a scalable technique for automated identification of explicit value-related dependencies and their characteristics (quality and strength) on the basis that values of software requirements are determined by user preferences for those requirements. In this regard, we have demonstrated using Eells measure of casual strength~\cite{eells1991probabilistic} for extracting qualities and strengths of value-related dependencies from user preferences. We have further, discussed value implications of certain structural and semantic dependencies among software requirements.

Finally, we have enhanced the accuracy of our proposed dependency identification technique through generating samples of large quantities based on the estimated distribution of the collected user preferences. In this regard, we have made use of a resampling~\cite{kroese2014statistical} technique that generates samples from user preferences using a Latent Multivariate Gaussian model~\cite{kroese2014statistical}. The employed resampling technique is computationally efficient and feasible for large numbers of software requirements~\cite{macke2009generating}.

We have verified the validity of the work by carrying out several simulations. Our results show that: (a) the DA-SRP model can properly consider qualities and strengths of value-related requirement dependencies in a SRP process while being computationally scalable, (b) The proposed DA-SRP model efficiently mitigates the adverse impact of the selection deficiency problem SDP, (c) the DA-SRP model always maximizes the overall value of the selected requirements, and (d) maximizing the overall value of a subset of requirements, where value-related dependencies are considered, and maximizing the \textit{Accumulated Value} (AV) of that subset, where value-related dependencies are not considered, are conflicting objectives~\cite{korte_combinatorial_2006}.


\section{Related work}
\label{sec_related}

Several works~\cite{carlshamre_industrial_2001,dahlstedt2005requirements,van_den_akker_flexible_2005, li_integrated_2010,sagrado_multi_objective_2013,Zhang_RIM_2013} have focused the attentions on considering value-related dependencies in software release planning (SRP). On this basis, we categorize the existing software release planning models into there main groups as follows. 

The first group~\cite{karlsson_optimizing_1997,jung_optimizing_1998,ruhe_trade_2003} of release planning models referred to as \textit{Binary Knapsack Problem} (BKP) models find a subset of requirements with the highest accumulated value while ignoring value-related dependencies among requirements. Such models hence are formulated as the classical binary knapsack problem~\cite{veerapen2015integer} as given by (\ref{Eq_BKP})-(\ref{Eq_BKP_c2}), where for a given set of requirements $R:\{r_1,...,r_n\}$, $c_i$ and $v_i$ denote the (estimated) cost and the (estimated) value of a requirement $r_i \in R$ respectively. Also, $b$ denotes the available budget and decision variable $x_i$ specifies whether a requirement $r_i$ is selected ($x_i=1$) or not ($x_i=0$). 

\small
\begin{align}
 \label{Eq_BKP}
  &\text{Maximize} \sum_{i=1}^{n} v_i x_i   \\
  \label{Eq_BKP_c1}
  &\text{Subject to} \sum_{i=1}^{n} c_i x_i \leq b\\
  \label{Eq_BKP_c2}
  & x_i \in \{0,1\},\quad i = 1,...,n
\end{align}
 \normalsize
 
The second group of software release planning models~\cite{li_integrated_2010,sagrado_multi_objective_2013,boschetti_lagrangian_2014,brasil_multiobjective_2012,bagnall_next_2001,Greer_evolutionary_2004,baker_search_2006} referred to as \textit{Binary Knapsack with Precedence Constraint} (BKP-PC) models, only capture value-related dependencies of full strength and formulate them as precedence constraints as given by (\ref{Eq_BKP-PC})-(\ref{Eq_BKP-PC_c3}). The rest of the value-related dependencies thus, have to be either ignored or treated as full-strength dependencies. Following the latter direction, a positive value-related dependency from a requirement $r_j$ to $r_k$ ($Q(r_j,r_k)=+$) is modeled as $x_j\le x_k$ while a negative value-related dependency ($Q(r_j,r_k)=-$) is modeled as $x_j\le 1-x_k$. $Q(r_j,r_k)$ gives the quality of the dependency from $r_j$ to $r_k$.  

\small
\begin{align}
\label{Eq_BKP-PC}
&\text{Maximize}  \sum_{i=1}^{n} v_i x_i   \\
\label{Eq_BKP-PC_c1}
&\text{Subject to} \sum_{i=1}^{n} c_i  x_i \leq b \\
\label{Eq_BKP-PC_c2}
& \begin{cases}
x_j \le x_k   & \text{$Q(r_j,r_k)=+$} \\
x_j \le 1-x_k & \text{$Q(r_j,r_k)=-\phantom{s} , j\neq k=1,...,n$} 
\end{cases}\\
\label{Eq_BKP-PC_c3}
& x_i \in \{0,1\},\hspace{5.65em} i=1,...,n 
\end{align}
\normalsize

Ignoring the strengths of value-related dependencies and molding them as precedence constraints however, results in a condition referred to as the selection deficiency problem (SDP)~\cite{mougouei2016factoring}. The reason is that when value-related dependencies are formulated as precedence constraints of (\ref{Eq_BKP-PC_c2}), ignoring (selecting) a requirement $r_i$ would result in ignoring all of the requirements whose values positively (negatively) depend on selection of $r_i$ even if the available budget is enough to allow for their implementation~\cite{li_integrated_2010}. Hence, employing BKP-PC models may result in loss of value due to the selection deficiency problem. The effect of the SDP is also demonstrated in a simulation study carried out by Li \textit{et al.}~\cite{li_integrated_2010} where a $2\%$ increase in the number of requirement dependencies resulted in almost $10\%$ loss of value in the optimal set.

To care for variances of value-related dependencies, the third group of release planning models~\cite{van_den_akker_flexible_2005,li_integrated_2010,sagrado_multi_objective_2013,Zhang_RIM_2013} referred to as \textit{Increase-Decrease} models consider value-related dependencies among software requirements through estimating the amount of the increased (decreased) values resulted by selecting various subsets of requirements. 

An Increases-Decreases model proposed by Akker \textit{et al.}~\cite{van_den_akker_flexible_2005} is given in (\ref{Eq_BKP-CS})-(\ref{Eq_BKP-CS_c3}). For a subset $ s_j \in S:\{s_1,...,s_m\}, m\leq2^n$, with $n_j$ requirements, the difference between the estimated value of $s_j$ ($w_j$) and the accumulated value of the requirements in $s_j$ ($\sum_{r_k\in s_j}v_k$) is considered when computing the value of the selected requirements. $y_j$ in (\ref{Eq_BKP-CS}) specifies whether a subset $s_j$ is realized ($y_j=1$) or not ($y_j=0$). Also, constraint~(\ref{Eq_BKP-CS_c2}) ensures that $y_j=1$ only if $\forall r_k \in s_j, x_k=1$.

\small
\begin{align}
\label{Eq_BKP-CS}
&\text{Maximize} \sum_{i=1}^{n} v_i x_i + \sum_{j=1}^{m}(w_j-\sum_{r_k\in s_j}v_k)\text{ } y_j\\
\label{Eq_BKP-CS_c1}
&\text{Subject to}\quad n_j y_j \leq \sum_{r_k \in s_j} x_k \\
\label{Eq_BKP-CS_c2}
&\sum_{i=1}^{n} c_i  x_i \leq b \\
\label{Eq_BKP-CS_c3}
& x_i,y_j \in \{0,1\} \quad i,j = 1,...,n
\end{align}
\normalsize

Increase-Decrease models however, are complex, subjective, and prone to human error as they rely on manual estimations for various subsets of requirements~\cite{mougouei2016factoring}. These problems persist even when estimations are limited to pairs of requirements~\cite{li_integrated_2010,sagrado_multi_objective_2013,Zhang_RIM_2013} as given in (\ref{Eq_BKP-Others})-(\ref{Eq_BKP-Others_c2}). For $n$ requirements, a pairwise approach would reduce the worst case complexity of manual estimations from estimating all subsets ($O(2^n)$) to estimating the subsets of size two only ($O(n^2)$) which is still way too complex to be used in software release planning. 

\begin{align}
\label{Eq_BKP-Others}
&\text{Maximize } \sum_{i=1}^{n} v_i x_i + \sum_{i=1}^{n}\sum_{j=1}^{n} w_{i,j} y_{i,j}\\
\label{Eq_BKP-Others_c1}
&\text{Subject to} \sum_{i=1}^{n} c_i  x_i \leq b \\
\label{Eq_BKP-Others_c2}
& x_i,y_{i,j} \in \{0,1\} \quad i,j = 1,...,n
\end{align}

\vspace{0.25cm}

Furthermore, relying on pairwise estimations results in ignoring implicit value-related dependencies among software requirements as the direction of dependencies are not specified in estimated increased (decreased) values. For instance, consider requirements $R:\{r_1,r_2,r_3\}$, where the value of $r_1$ positively depends on $r_2$ and the value of $r_2$ positively depends on $r_3$. On this basis, it is logical to infer an implicit positive value-related dependency from $r_1$ to $r_3$. An Increase-Decrease model however, fails to capture this implicit dependency even if pairwise estimations show that selecting $r_1$ and $r_2$ as a pair, increases the value of that pair and the estimated value of the requirements $r_2$ and $r_3$ increases when selected together. As such, if no explicit value-related dependency is identified between $r_1$ and $r_3$ by pairwise estimations, the influence of $r_3$ on the value of $r_1$ will be ignored by the Increase-Decrease model.

\section{Modeling Value-Related Dependencies by Fuzzy Graphs}
\label{sec_modeling}

It is widely known that using graphs for modeling requirement dependencies~\cite{carlshamre_industrial_2001,bohner1996software,wilde1991reusable,ngo_fuzzy_2005,hapke1994fuzzy} results in enhancing the accuracy of release planning~\cite{carlshamre_industrial_2001,mougouei2017modeling}. On the other hand, fuzzy graphs and their algebraic structure have demonstrated to contribute to more accurate models as they can properly capture the imprecision associated with requirement dependencies~\cite{ngo2005measuring,wang_simulation_2012}.

On this basis, we have introduced Value Dependency Graphs (VDGs) based on the classical fuzzy graphs for modeling value-related dependencies and their characteristics (quality and strength) in software projects. We have specially modified the classical definition of fuzzy graphs to consider not only the strength but also the quality of value-related dependencies as given by Definition~\ref{def_vdg}. 

\begin{defn}
\label{def_vdg}
\textit{Value Dependency Graph (VDG)}. A VDG is a signed directed fuzzy graph~\cite{Wasserman1994} $G=(R,\sigma,\rho)$ in which a non-empty set of identified requirements $R:\{r_1,...,r_n\}$ constitute the graph nodes. The qualitative function $\sigma: R\times R\rightarrow \{+,-,\pm\}$ and the membership function $\rho: R\times R\rightarrow [0,1]$ denote qualities (positivity and negativity) and strengths of explicit value-related dependencies (edges of the graph) receptively. As such, a pair of requirements $(r_i,r_j)$ with $\rho(r_i,r_j)\neq 0$ denotes a positive (negative) value-related dependency from $r_i$ to $r_j$ stating that selection of $r_j$ has an explicit positive (negative) influence on the value of $r_i$ which is equivalent to $r_j$ is a positive (negative) value-related dependence of $r_i$. 
\end{defn}

\begin{figure}[!htb]
	\begin{center}
		\includegraphics[scale=0.6]{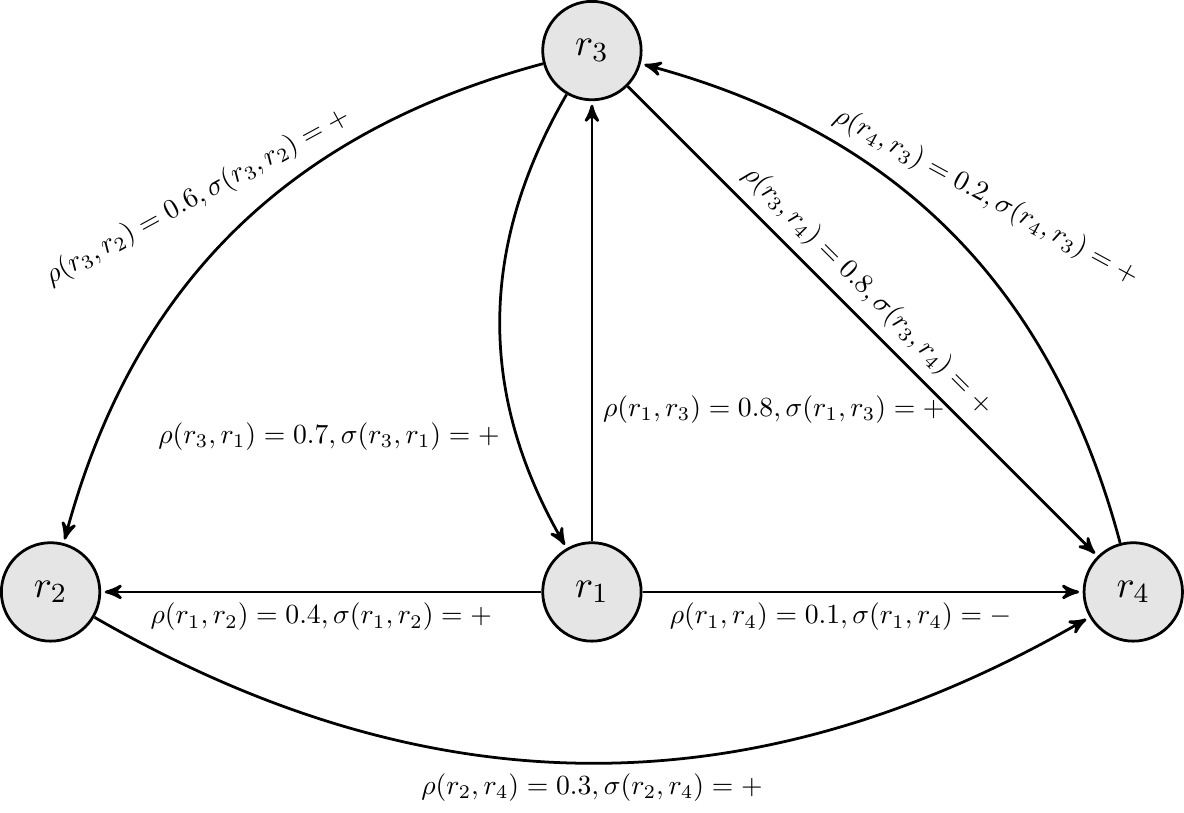}
	\end{center}
	\caption{%
		A Sample Value Dependency Graph.}%
	\label{fig_ex_vdg}
\end{figure}

For instance, in the VDG $G=(R,\sigma,\rho)$ of Figure~\ref{fig_ex_vdg} with $R=\{r_1,r_2,r_3,r_4,r_5\}$, $\sigma(r_1,r_2)=+$ and $\rho(r_1,r_2)=0.4$ state that selection of requirement $r_2$ has an explicit positive influence on the value of requirement $r_1$ and the strength of this influence is roughly $0.4$. It is clear that we have $\rho(r_i,r_j)=0$ if the value of a requirement $r_i$ is not explicitly influenced by selection of $r_j$. In such cases we have $\sigma(r_i,r_j)=\pm$, where $\pm$ denotes the quality of $(r_i,r_j)$ is \text{non-specified}. 


\begin{defn}
	\label{def_vdg_valuedepndencies}
	\textit{Value-Related Dependencies}. 
	A value-related dependency in a VDG $G=(R,\sigma,\rho)$ is defined as a sequence of requirements $d_i:\big(r(0),...,r(k)\big)$ such that $\forall r(j) \in d_i$, $1 \leq j \leq k$ we have $\rho\big(r(j-1),r(j)\big) \neq 0$. $j\geq 0$ is the sequence of the $j^{th}$ requirement (node) denoted as $r(j)$ on the dependency path. A consecutive pair $\big(r(j-1),r(j)\big)$ specifies an explicit value-related dependency. 
\end{defn}

The strength of a value-related dependency $d_i:\big(r(0),...,r(k)\big)$ is calculated by (\ref{Eq_vdg_strength}). That is, the strength of a value-related dependency equals to the strength of the weakest of all the $k$ explicit value-related dependencies along the path. Fuzzy operator $\wedge$ denotes Zadeh's~\cite{zadeh_fuzzysets_1965} fuzzy AND operation (taking infimum). 

\small
\begin{align}
\label{Eq_vdg_strength}
\forall d_i:\big(r(0),...,r(k)\big): &\rho(d_i) = \\\nonumber 
& \bigwedge_{j=1}^{k}\text{ }\rho\big(r(j-1),r(j)\big) 
\end{align}

\begin{align}
\label{Eq_vdg_quality}
\forall d_i:\big(r(0),...,r(k)\big): &\sigma(d_i) = \\\nonumber
& \prod_{j=1}^{k}\text{ }\sigma\big(r(j-1),r(j)\big)
\end{align}

\normalsize

On the other hand, the quality (positive/negative) of a value-related dependency $d_i:\big(r(0),...,r(k)\big)$ is calculated through qualitative serial inference~\cite{de1984qualitative,wellman1990formulation,kusiak_1995_dependency} as given by (\ref{Eq_vdg_quality}) and Table~\ref{table_inference}. For instance, given $d_i:\big(r(0),r(1),r(2)\big)$, we can infer from $ \sigma\big(r(0),r(1)\big)=+ $ and $ \sigma\big(r(1),r(2)\big)=-$ that $\sigma\big(r(0),r(1),r(2)\big)=- $. This can be informally proved as given in (\ref{Eq_serial_inference}), where $\neg r(2)$ denotes ignoring $r(2)$.

\begin{table}[!htb]
	\caption{Qualitative Serial Inference in VDGs.}
	\label{table_inference}
	\centering
\resizebox {0.3\textwidth }{!}{
	\begin{tabular}{cc|ccc}
		\toprule[1.5pt]
		\multicolumn{2}{r|}{\multirow{2}[1]{*}{ $\sigma\big(r(j-1),r(j),r(j+1)\big)$}} &
		\multicolumn{3}{c}{$\sigma\big(r(j),r(j+1)\big)$}
		\\
		\multicolumn{2}{r|}{} &
		$+$ &
		$-$ &
		$\pm$
		\bigstrut[b]\\
		\hline
		\multicolumn{1}{c}{\multirow{3}[1]{*}{$\sigma\big(r(j-1),r(j)\big)$}} &
		$+$ &
		$+$ &
		$-$ &
		$\pm$
		\bigstrut[t]\\
		\multicolumn{1}{c}{} &
		$-$ &
		$-$ &
		$+$ &
		$\pm$
		\\
		\multicolumn{1}{c}{} &
		$\pm$ &
		$\pm$ &
		$\pm$ &
		$\pm$
		\\
    \bottomrule[1.5pt]
	\end{tabular}%
	}

\end{table}

\small
\begin{align}
\label{Eq_serial_inference}
 &(a):\quad \sigma\big(r(0),r(1)\big)=+ \\ \nonumber
 &(b):\quad \sigma\big(r(1),r(2)\big)=-\rightarrow \sigma\big(r(1),\neg r(2)\big)=+ \\ \nonumber
 &(a),(b)\Rightarrow \sigma\big(r(0),r(1),\neg r(2)\big)= + 
 \\\nonumber 
 &\rightarrow \sigma\big(r(0),r(1),r(2)\big)=-
\end{align}
\normalsize

Quantitative serial inferences in Table~\ref{table_inference} are mathematically proved by Wellman~\cite{wellman1990formulation} and Kleer~\cite{de1984qualitative}. Also, Kusiak~\cite{kusiak_1995_dependency} has explained the details of employing them in constraint negotiation.
 

Let $D=\{d_1,d_2,..., d_m\}$ be the set of all explicit and implicit value-related dependencies from $r_i \in R$ to $r_j \in R$ in a VDG $G=(R,\sigma,\rho)$. As explained earlier, positive and negative value-related dependencies can simultaneously exist from $r_i$ to $r_j$. The strength of all positive value-related dependencies from $r_i$ to $r_j$ is denoted by $\rho^{+\infty}(r_i,r_j)$ and calculated by (\ref{Eq_ultimate_strength_positive}), that is to find the strength of the strongest positive value-related dependency~\cite{rosenfeld_fuzzygraph_1975} among all the positive value-related dependencies from $r_i$ to $r_j$. Fuzzy operators $\wedge$ and $\vee$ denote Zadeh's~\cite{zadeh_fuzzysets_1965} fuzzy AND (taking minimum over AND-ed operands) and fuzzy OR (taking maximum over OR-ed operands) operations respectively. In a similar way, the strength of all negative value-related dependencies from $r_i$ to $r_j$ is denoted by $\rho^{-\infty}(r_i,r_j)$ and calculated by (\ref{Eq_ultimate_strength_negative}).

\small
\begin{align}
\label{Eq_ultimate_strength_positive}
&\rho^{+\infty}(r_i,r_j) = \bigvee_{d_m\in D, \sigma(d_m)=+} \text{ } \rho(d_m) \\[2pt]
\label{Eq_ultimate_strength_negative}
&\rho^{-\infty}(r_i,r_j) = \bigvee_{d_m\in D, \sigma(d_i)=-} \text{ } \rho(d_m) 
\end{align}
\normalsize
\vspace{0.25cm}

A brute-force approach to calculating $\rho^{+\infty}(r_i,r_j)$ or $\rho^{-\infty}(r_i,r_j)$ would compute the strengths of all possible paths from $r_i$ to $r_j$ which could get as complex as $O(n!)$, where $n$ denotes the number of requirements (VDG nodes). To avoid such complexity, we have formulated the problem of calculating $\rho^{+\infty}(r_i,r_j)$ and $\rho^{-\infty}(r_i,r_j)$ as a widest path problem (also known as the maximum capacity path problem~\cite{vassilevska_all_2007}) which can be solved in polynomial time by Floyd-Warshall~\cite{floyd_1962} algorithm. 

In this regard, we devised a modified version of Floyd-Warshall algorithm (Algorithm~\ref{alg_strength}) that computes $\rho^{+\infty}(r_i,r_j)$ and $\rho^{-\infty}(r_i,r_j)$ for all pairs of requirements: $\forall (r_i,r_j), r_i,r_j \in R:\{r_1,...,r_n\}$ with the time bound of $O(n^3)$. For each pair of requirements $(r_i,r_j)$ in a VDG $G=(R,\sigma,\rho)$, lines $20$ to $36$ of Algorithm~\ref{alg_strength} find the strength of all positive value-related dependencies and the strength of all negative value-related dependencies from $r_i$ to $r_j$.

\begin{algorithm*}
    \small
	\caption{Calculating Strengths of Value Dependencies.}
	\label{alg_strength}
	\begin{multicols}{2}
		\begin{algorithmic}[1]
					\REQUIRE VDG $G=(R,\sigma,\rho)$
					\ENSURE $\rho^{+\infty}, \rho^{-\infty}$
					\FOR{\textbf{each} $r_i \in R$}
					\FOR{\textbf{each} $r_j \in R$}
					\STATE $\rho^{+\infty}(r_i,r_j) \leftarrow \rho^{-\infty}(r_i,r_j) \leftarrow -\infty$ 
					\ENDFOR
					\ENDFOR
					\FOR{\textbf{each} $r_i \in R$}
					\STATE $\rho(r_i,r_i)^{+\infty} \rho(r_i,r_i)^{-\infty} \leftarrow 0$
					\ENDFOR
					\FOR{\textbf{each} $r_i \in R$}
					\FOR{\textbf{each} $r_j \in R$}
					\IF{$\sigma(r_i,r_j) = +$}
					\STATE $\rho^{+\infty}(r_i,r_j) \leftarrow \rho(r_i,r_j)$
					\ELSIF{$\sigma(r_i,r_j) = -$}
					\STATE $\rho^{-\infty}(r_i,r_j) \leftarrow \rho(r_i,r_j)$
					\ENDIF
					\ENDFOR
					\ENDFOR
					\FOR{\textbf{each} $r_k \in R$}
					\FOR{\textbf{each} $r_i \in R$}
					\FOR{\textbf{each} $r_j \in R$}
					\IF{$min\big(\rho(r_i,r_k)^{+\infty}, \rho(r_k,r_j)^{+\infty}\big) > \rho^{+\infty}(r_i,r_j)$}
					\STATE $\rho^{+\infty}(r_i,r_j) \leftarrow  min(\rho(r_i,r_k)^{+\infty}, \rho(r_k,r_j)^{+\infty})$
					\ENDIF
					\IF{$min\big(\rho(r_i,r_k)^{-\infty}, \rho(r_k,r_j)^{-\infty}\big) > \rho^{+\infty}(r_i,r_j)$}
					\STATE $\rho^{+\infty}(r_i,r_j) \leftarrow  min(\rho(r_i,r_k)^{-\infty}, \rho(r_k,r_j)^{-\infty})$
					\ENDIF
					\IF{$min\big(\rho(r_i,r_k)^{+\infty}, \rho(r_k,r_j)^{-\infty}\big) > \rho^{-\infty}(r_i,r_j)$}
					\STATE $\rho^{-\infty}(r_i,r_j) \leftarrow  min(\rho(r_i,r_k)^{+\infty}, \rho(r_k,r_j)^{-\infty})$
					\ENDIF
					\IF{$min\big(\rho(r_i,r_k)^{-\infty}, \rho(r_k,r_j)^{+\infty}\big) > \rho^{-\infty}(r_i,r_j)$}
					\STATE $\rho^{-\infty}(r_i,r_j) \leftarrow  min(\rho(r_i,r_k)^{-\infty}, \rho(r_k,r_j)^{+\infty})$
					\ENDIF
					\ENDFOR
					\ENDFOR
					\ENDFOR
		\end{algorithmic}
	\end{multicols}
\end{algorithm*}

The overall strength of all positive and negative value-related dependencies from $r_i$ to $r_j$ is referred to as the \textit{Overall Influence} of $r_j$ on the value of $r_i$ and denoted by $I_{i,j}$. $I_{i,j}$ as given by (\ref{Eq_influence}) is calculated by subtracting the strength of all negative value-related dependencies from $r_i$ to $r_j$ ($\rho(r_i,r_j)^-\infty$) from the strength of all positive value-related dependencies from $r_i$ to $r_j$ ($\rho(r_i,r_j)^+\infty$). It is clear that $I_{i,j}\in[-1,1]$. $I_{i,j}>0$ states that $r_j$ influences the value of $r_i$ in a positive way whereas $I_{i,j}<0$ indicates that the ultimate influence of $r_j$ on $r_i$ is negative.  

\small
\begin{align}
\label{Eq_influence}
I_{i,j} = \rho^{+\infty}(r_i,r_j)-\rho^{-\infty}(r_i,r_j) 
\end{align}
\normalsize
%
%

\begin{exmp}
	\label{ex_ultimate_strength}
	Consider the set of all value-related dependencies from requirement $r_1$ to requierment $r_4$: $D=\{d_1=(r_1,r_2,r_4),d_2=(r_1,r_3,r_4),d_3=(r_1,r_4)\}$ in the VDG of Figure \ref{fig_ex_vdg}. Using (\ref{Eq_vdg_quality}) qualities of value-related dependencies in $D$ are calculated as follows. $\sigma(d_1)=\Pi(+,+)=+$, $\sigma(d_2)=\Pi(+,+)=+$, and $\sigma(d_3)=\Pi(-)=-$. Also strengths of value-related dependencies in $D$ are calculated by (\ref{Eq_vdg_strength}) as : $\rho(d_1)=\wedge\big(\rho(r_1,r_2),\rho(r_2,r_4)\big)=min(0.4,0.3)=0.3$, $\rho(d_2)=\wedge\big(\rho(r_1,r_3),\rho(r_3,r_4)\big)=min(0.8,0.8)=0.8$,  $\rho(d_3)=0.1$. Using~(\ref{Eq_ultimate_strength_positive}) we have: $\rho(r_1,r_4)^+\infty = \vee(\rho(d_1),\rho(d_2))=max(0.3,0.8)=0.8$. Also using~(\ref{Eq_ultimate_strength_negative}) we have $\rho^{-\infty}(r_1,r_4) = \rho(d_3)=0.1$. Therefore, we have $I_{1,4} = \rho(r_1,r_4)^{+\infty}-\rho(r_1,r_4)^{-\infty}=0.7$. $I_{1,4}=0.7 > 0$ indicates that positive influence of $r_4$ on the value of $r_1$ prevails. Table~\ref{table_ex_overall} lists overall influences of requirements of VDG of Figure~\ref{fig_ex_vdg} on the values of each other.
\end{exmp}
\begin{table*}[!htbp]
	\centering
	\caption{Overall influences for VDG of Figure~\ref{fig_ex_vdg}.}
	\label{table_ex_overall}
\resizebox {0.8\textwidth }{!}{
	\begin{tabular}{lcccc}
		\toprule[1.5pt]
		\textbf{\cellcolor{black}\textcolor{white}{$I_{i,j}=\rho(r_i,r_j)^{+\infty}-\rho(r_i,r_j)^{-\infty}$}}&
		\textbf{\cellcolor{black}\textcolor{white}{$r_1$}}&
		\textbf{\cellcolor{black}\textcolor{white}{$r_2$}}&
		\textbf{\cellcolor{black}\textcolor{white}{$r_3$}}&
		\textbf{\cellcolor{black}\textcolor{white}{$r_4$}}
		\\ \midrule
		\textbf{$r_1$}\unboldmath{} &
		$0.0-0.0=0.0$ &
		$0.6-0.1=0.5$ &
		$0.8-0.1=0.7$ &
		$0.8-0.1=0.7$
		\\
		\textbf{$r_2$}\unboldmath{} &
		$0.2-0.0=0.2$ &
		$0.0-0.0=0.0$ &
		$0.2-0.0=0.2$ &
		$0.3-0.0=0.3$
		\\
		\textbf{$r_3$}\unboldmath{} &
		$0.7-0.1=0.6$ &
		$0.6-0.1=0.5$ &
		$0.0-0.0=0.0$ &
		$0.8-0.1=0.7$
		\\
		\textbf{$r_4$}\unboldmath{} &
		$0.2-0.0=0.2$ &
		$0.2-0.0=0.2$ &
		$0.2-0.0=0.2$ &
		$0.0-0.0=0.0$
		\\ \bottomrule[1.5pt]
	\end{tabular}
	}

\end{table*}
\begin{defn}
	\label{def_frig_VDL}
	\textit{Value Dependency Level (VDL) and Negative Value Dependency Level (NVDL)}. Let $G=(R,\sigma,\rho)$ be a VDG with $R=\{r_1,...,r_n\}$, $k$ be the total number of explicit value-related dependencies in $G$, and $m$ be the total number of negative explicit value-related dependencies. Then, VDL and NVDL of $G$ are derived by (\ref{Eq_vdl}) and (\ref{Eq_nvdl}) respectively. 
\end{defn}
\small
\begin{align}
\label{Eq_vdl}
&VDL(G)=\frac{k}{\Perm{n}{2}}=\frac{k}{n (n-1)} \\
\label{Eq_nvdl}
&NVDL(G)=\frac{m}{k}
\end{align}
\normalsize
\vspace{0.25cm}
\begin{exmp}
	\label{ex_vdl}
	Consider the value-related dependency graph $G$ of Figure~\ref{fig_ex_vdg} for which we have $n=4$, $k=8$, and $m=1$. $VDL(G)$ is derived by~(\ref{Eq_vdl}) as: $VDL(G)= \frac{8}{4\times 3} = \frac{8}{12} \approxeq 0.67$. Also we have from Equation~(\ref{Eq_nvdl}), $NVDL(G)=\frac{1}{8}=0.125$.
\end{exmp}
\vspace{0.25cm}
\section{Identifying Explicit Value-Related Dependencies}
\label{sec_identification}

It has been widely known that the value of a software requirement is determined by user preferences (satisfaction) of that requirement, that applies to monetary as well as nonmonetary definitions of value~\cite{mougouei2018operationalizing,perera2019study,perera2019towards,hussain2018integrating}. Hence, terms like \textit{customer value} or \textit{business value} have been preferred by some of the existing works~\cite{racheva2010business,villarroel2016release} to emphasize the importance of user (customer) preferences in specifying the values of software requirements. User preferences for a software requirement however, may be influenced by user preferences for other requirements. In other words, users preferring (selecting or using) a requirement $r_j$ maybe be more likely or less likely to prefer a requirement $r_i$. As such, the (customer) value of a software requirement can be influenced by user preferences for other requirements. Based on this logic, value-related dependencies among software requirements can be inferred from causal relations~\cite{sprenger2016foundations,Halpern01062015,pearl2009causality,janzing2013quantifying,eells1991probabilistic} among user preferences for those requirements. For instance, when user preferences for a requirement $r_j$ positively influences user preferences for a requirement $r_i$, this implies a value-related dependency from $r_i$ to $r_j$ with $\rho(r_i,r_j)\neq 0$ and $\sigma(r_i,r_j)>0$. 
%
%

User preferences can be gathered in different ways~\cite{villarroel2016release,leung2011probabilistic,holland2003preference,sayyad2013value} depending on the nature of a software release. For the very first release of a software, user preferences can be gathered by conventional market research approaches such as conducting surveys or referring to the user feedbacks or sales records of the similar software products in the market. For future releases of a software, or when re-engineering of a software is of interest (e.g. for legacy systems) however, user feedbacks and sales records of the previous releases of a software might be used in combination with market research results to find user preferences. Once user preferences are gathered, we can identify explicit value-related dependencies and their characteristics (namely quality and strength) using measures of causal strength~\cite{sprenger2016foundations,Halpern01062015,pearl2009causality,janzing2013quantifying,eells1991probabilistic}. 
%

In this paper, we have used one of the most widely adopted measures of causal strength referred ton as the Eells measure ~\cite{eells1991probabilistic}. Eells measure for an explicit value-related dependency $(r_i,r_j)$ is denoted by $\eta_{i,j}$ and computed by (\ref{Eq_Eells}). $\eta_{i,j}$ specifies the extent to which selecting or ignoring $r_j$ influences user preferences of $r_i$ and consequently the value of $r_i$. Eells measure properly captures both positive and negative value-related dependencies between a pair of requirements through subtracting the conditional probability $p(r_i|\neg r_j)$ from $p(r_i|r_j)$, where conditional probabilities $p(r_i|\neg r_j)$ and $p(r_i|r_j)$ denote strengths of positive and negative dependencies from $r_i$ to $r_j$ respectively. $\neg r_j$ denotes ignoring $r_j$. 

\vspace{-0.25cm}

\small
\begin{align}
\label{Eq_Eells}
& \eta_{i,j}= p(r_i|r_j) - p(r_i|\neg r_j) ,\phantom{s}\eta_{i,j} \in [-1,1]
\end{align}
\normalsize

Complexity of computing Eells measure for all pairs of requirements in a requirement set of size $n$ is $n^2 \times u$, where $u$ is the number of users whose preferences are gathered. As such, the process is scalable to large number of requirements. Hence, automated identification of value-related dependencies adds a negligible overhead to software release planning. 

\small
\begin{align}
\label{Eq_qualityMeasure}
& \sigma(r_i,r_j) =  \begin{cases}
+ & \text{if }\phantom{s}  \eta_{i,j} > 0 \\
- & \text{if }\phantom{s}  \eta_{i,j} < 0 \\
\pm & \text{if }\phantom{s} \eta_{i,j} = 0 \\
\end{cases}
\end{align}
\normalsize

To identify quality of a value-related dependency $(r_i,r_j)$, (\ref{Eq_qualityMeasure}) gives a mapping from $\eta_{i,j}$ to the qualitative function $\sigma: R\times R\rightarrow \{+,-,\pm\}$, where $\eta_{i,j}>0$ indicates that the strength of the positive dependency from $r_i$ to $r_j$ is greater than the strength of its corresponding negative dependency: $p(r_i|r_j) > p(r_i|\neg r_j)$ and therefore the quality of $d$ is positive ($\sigma(r_i,r_j)=+$). Similarly, $\eta_{i,j}<0$ indicates $p(r_i|\neg r_j) > p(r_i|r_j)$ which means $\sigma(r_i,r_j)=-$. Also, $p(r_i|r_j) - p(r_i|\neg r_j)=0$ specifies that the quality of the zero-strength value-related dependency $(r_i,r_j)$ is non-specified ($\sigma(r_i,r_j)=\pm$).

\begin{figure}[!htb]
	\begin{center}
		\subfigure[$$]{%
			\label{fig_membership_1}
			\includegraphics[scale=0.71]{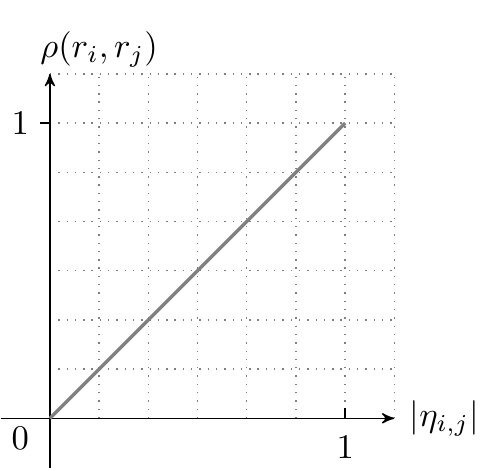}
		}
		\subfigure[$$]{%
			\label{fig_membership_2}
			\includegraphics[scale=0.71]{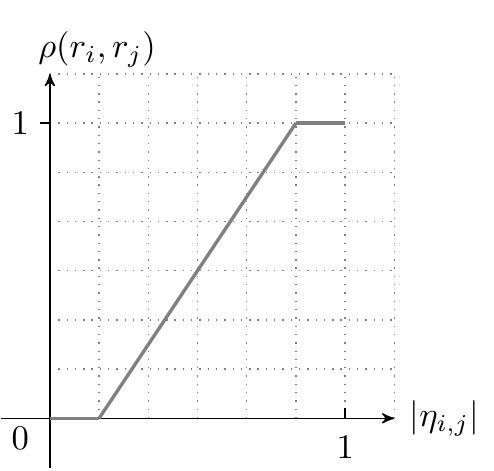}
		}
	\end{center}
	\vspace{-0.3cm}
	\caption{%
		Sample membership functions $\rho(r_i,r_j)$.
	}%
	\label{fig_membership}
\end{figure}

To compute the strength of a value-related dependency $(r_i,r_j)$ in a VDG $G=(R,\sigma,\rho)$, (\ref{Eq_strengthMeasure}) gives a mapping from $\eta_{i,j}$ to the fuzzy membership function $\rho: R\times R\rightarrow [0,1]$. This mapping is demonstrated in Figure~\ref{fig_membership_1}. Nevertheless, other membership functions such as the one in Figure~\ref{fig_membership_2} could also be used for this mapping~\cite{mougoueifuzzy2019} - depending on the desired behavior of the release planning model used. For instance, the membership function of Figure~\ref{fig_membership_1} ignores value-related dependencies with casual strengths below $0.16$ ($\eta_{i,j} < 0.16$) while  dependencies with $\eta_{i,j} \geq 0.83$ are considered strong enough to be formulated as full strength dependencies ($\rho(r_i,r_j)=1$). 

\begin{align}
\label{Eq_strengthMeasure}
& \rho(r_i,r_j)= |\eta_{i,j}|
\end{align}


\vspace{0.5cm}
When the number of user preferences gathered is limited, resampling can be used to generate samples of large (enough) quantities based on the estimated distribution of the collected user preferences. To achieve this, we use a scalable technique proposed by Macke \textit{et al.}~\cite{macke2009generating} to generate larger samples of user preferences using a Latent Multivariate Gaussian model~\cite{kroese2014statistical}. Our resampling process starts with computing estimated means and variances of user preferences for each requirement. Then, covariances matrix of the requirements will be computed for generating new samples. Finally, samples of desired size will be generated based on the Dichotomized Gaussian Distribution model discussed in~\cite{kroese2014statistical}.

Finally, it is worth mentioning that, \textit{Intrinsic} (structural or semantic) dependencies of full strength among software requirements (e.g. \textit{requires}~\cite{dahlstedt2005requirements} and \textit{conflicts with}~\cite{k_process_centered_1996}) also have value implications. For instance, a requirement $r_i$ requires (conflicts with) $r_j$ means that $r_i$ can not give any value if $r_j$ is ignored (selected). Hence, it is also important to be aware of value implications of intrinsic dependencies and consider them in software release planning. This could be achieved by carefully studying the structure (semantic) of a software.  


\section{The Integer Programming Model}
\label{sec_factoring}

As explained earlier, the value of a software requirement is determined by user preferences for that requirement~\cite{zhang_investigating_2014,biffl_value_2006,racheva2010business}. In other words, the higher the ratio of the users that prefer (select or use) a software requirement $r_i$, the higher the expected value of $r_i$ would be. That means $r_i$ can achieve its highest expected value (which equals to its estimated value) when all users of $r_i$ are satisfied with that requirement. However, this only happens when user preferences for the requirement $r_i$ is not negatively influenced by selecting or ignoring other requirements. In other cases, a reduction from the estimated value of $r_i$ is predictable. The extent of such reduction is referred to as the penalty~\cite{wiegers_software_2009,mougouei2017bkp} of $r_i$ and denoted as $p_i$. 

\small
\begin{align}
\label{Eq_penalty}
\nonumber
p_{i}= &\displaystyle \bigvee_{j=1}^{n} \bigg(\frac{x_j\big(\lvert I_{i,j} \rvert-I_{i,j}\big) + (1-x_j)\big(\lvert I_{i,j}\rvert+I_{i,j}\big)}{2}\bigg)=\\ 
&\displaystyle \bigvee_{j=1}^{n} \bigg(\frac{\lvert I_{i,j} \rvert + (1-2x_j)I_{i,j}}{2}\bigg),\quad i=1,...,n \\
& x_j \in\{0,1\},\quad \quad \quad \quad \quad \quad \quad  \hspace{0.2em}  j=1,...,n 
\end{align}
\normalsize

We have made use of the algebraic structure of fuzzy graphs for modeling value-related dependencies and their characteristics (quality and strength) as explained in Section~\ref{sec_modeling}. Hence, $p_i$ is calculated using fuzzy OR operator, which is to take supremum over the strengths of all ignored positive dependencies and selected negative dependencies of $r_i$ in its corresponding value dependency graph. This is given in (\ref{Eq_penalty}) and (\ref{Eq_dars_c2}), where $n$ denotes the number of requirements and $x_j$ specifies whether a requirement $r_j$ is selected ($x_j=1$) or not ($x_j=0$). Also, $I_{i,j}$ denotes the overall strength of all positive and negative value-related dependencies from $r_i$ to $r_j$ which is referred to as the overall influence of $r_j$ on the value of $r_i$ as given in (\ref{Eq_influence}). 

Equation (\ref{Eq_value}) derives the expected value of a software requirement $r_i$ denoted by $v^\prime_i$ which captures the impact of user preferences for $r_i$ ($\phi_i$) on the value of $r_i$. $\phi_i$ is derived by subtracting the penalty $p_i$ from the ideal user preference (satisfaction) level $1$ (when every user selects or uses $r_i$). As such, the overall value (OV) of a selected subset of requirements can be calculated by accumulating the expected values of the selected requirements as given by (\ref{Eq_ocv}), where $x_i$ denotes whether $r_i$ is selected ($x_i=1$) or not ($x_i=0$).

\small
\begin{align}
\label{Eq_cs}
& \phi_i = 1-p_i\\
\label{Eq_value}
& v^\prime_i = \phi_iv_i \\
\label{Eq_ocv}
&OV = \sum_{i=1}^{n} x_i (1-p_i)v_i, \textit{ } x_i \in \{0,1\}
\end{align}
\normalsize

\begin{exmp}
	\label{ex_overall}
	For a candidate set $s_1=\{r_1,r_2,r_3\}$ ($x_1=x_2=x_3=1$, $x_4=0$) in VDG of Figure \ref{fig_ex_vdg} consider finding penalties of selected requirements. From Table~\ref{table_ex_overall} we have $I_{1,4}=I_{3,4}=0.7,I_{2,4}=0.3,I_{4,4}=0.0$. As such, based on~(\ref{Eq_penalty}) penalties can be calculated as $p_{1}= \vee(\frac{\lvert 0.0 \rvert +(1-2(1))(0.0)}{2}$, $\frac{\lvert 0.45 \rvert +(1-2(1))(0.5)}{2}$, $\frac{\lvert 0.7 \rvert +(1-2(1))(0.7)}{2}$, $\frac{\lvert 0.7 \rvert +(1-2(0))(0.7)}{2}) =0.7$. In a similar way, we have $p_2=0.3,p_3=0.7$. By using~(\ref{Eq_cs}) we have $\phi_1=1-0.7=0.3$, $\phi_2=1-0.3=0.7,\phi_3=1-0.7=0.3$. Therefore, achieved expected values of selected requirements $r_1,r_2,r_3$ can be derived by~(\ref{Eq_value}) as: $v^\prime_1=0.3(20)=6, v^\prime_2 = 0.7(10)=7, v^\prime_3 = 0.3(50)=15$. Hence, we have $OV(s_1) = \sum_{i=1}^{4} x_iv^\prime_i= 28$.
\end{exmp}

Our proposed release planning model referred to as the DA-SRP (Dependency-Aware Software Release Planning), maximizes the overall value (OV) of a selected subset of requirements considering value-related dependencies among requirements. Equations (\ref{Eq_dars})-(\ref{Eq_dars_c4}) give the integer programming model for DA-SRP, where $x_i$ is a selection variable denoting whether a requirements $r_i$ is selected ($x_i=1$) or ignored ($x_i=0$). Also, $p_i$ specifies the penalty of a requirement $r_i$ which is the extent to which ignoring positive value-related dependencies and selecting negative value-related dependencies of $r_i$ reduce its expected value. 


\small
\begin{align}
\label{Eq_dars}
& \text{Maximize } \sum_{i=1}^{n} x_i (1-p_i) v_i\\
\label{Eq_dars_c1}
& \text{Subject to} \sum_{i=1}^{n} c_i x_i \leq b\\ 
\label{Eq_dars_c2}
& p_i \ge \displaystyle \bigg(\frac{\lvert I_{i,j} \rvert + (1-2x_j)I_{i,j}}{2}\bigg),& i\neq j = 1,...,n\\
\label{Eq_dars_c3}
&\text{ }x_i \in \{0,1\}& i = 1,...,n \\
\label{Eq_dars_c4}
&\text{ } 0 \leq p_i \leq 1& i = 1,...,n
\end{align}
\normalsize

\section{Simulations}
\label{sec_simulation}

\subsection{Simulation Design}
\label{simulation_design}


Simulation of requirement dependencies has been proposed in several works from the literature~\cite{li_integrated_2010,wang_simulation_2012,cai_evolutionary_2012} for analyzing the efficiency of release planning models. On this basis, we carried out simulations to compare efficiency of the DA-SRP model against that of the existing release planning models (BKP and BKP-PC models). The Increase-Decrease models however, could not be simulated as they do not specify how to formally achieve the amount of the increased or decreased values as explained in Section~\ref{sec_related}.

\begin{table*}[!htbp]
	\caption{Estimated costs and values of the requirements.}
	\label{table_cost_value}
	\centering
	\huge\resizebox {1\textwidth }{!}{
\begin{tabular}{lllllllllllllllllllllllllllll}
\toprule[1.5pt]
\textbf{\cellcolor{black}\textcolor{white}{Requirement}} &\textbf{\cellcolor{black}\textcolor{white}{$r_{1}$}}&
\textbf{\cellcolor{black}\textcolor{white}{$r_{2}$}}&
\textbf{\cellcolor{black}\textcolor{white}{$r_{3}$}}&
\textbf{\cellcolor{black}\textcolor{white}{$r_{4}$}}&
\textbf{\cellcolor{black}\textcolor{white}{$r_{5}$}}&
\textbf{\cellcolor{black}\textcolor{white}{$r_{6}$}}&
\textbf{\cellcolor{black}\textcolor{white}{$r_{7}$}}&
\textbf{\cellcolor{black}\textcolor{white}{$r_{8}$}}&
\textbf{\cellcolor{black}\textcolor{white}{$r_{9}$}}&
\textbf{\cellcolor{black}\textcolor{white}{$r_{10}$}}&
\textbf{\cellcolor{black}\textcolor{white}{$r_{11}$}}&
\textbf{\cellcolor{black}\textcolor{white}{$r_{12}$}}&
\textbf{\cellcolor{black}\textcolor{white}{$r_{13}$}}&
\textbf{\cellcolor{black}\textcolor{white}{$r_{14}$}}&
\textbf{\cellcolor{black}\textcolor{white}{$r_{15}$}}&
\textbf{\cellcolor{black}\textcolor{white}{$r_{16}$}}&
\textbf{\cellcolor{black}\textcolor{white}{$r_{17}$}}&
\textbf{\cellcolor{black}\textcolor{white}{$r_{18}$}}&
\textbf{\cellcolor{black}\textcolor{white}{$r_{19}$}}&
\textbf{\cellcolor{black}\textcolor{white}{$r_{20}$}}&
\textbf{\cellcolor{black}\textcolor{white}{$r_{21}$}}&
\textbf{\cellcolor{black}\textcolor{white}{$r_{22}$}}&
\textbf{\cellcolor{black}\textcolor{white}{$r_{23}$}}&
\textbf{\cellcolor{black}\textcolor{white}{$r_{24}$}}&
\textbf{\cellcolor{black}\textcolor{white}{$r_{25}$}}&
\textbf{\cellcolor{black}\textcolor{white}{$r_{26}$}}&
\textbf{\cellcolor{black}\textcolor{white}{$r_{27}$}}&
\textbf{\cellcolor{black}\textcolor{white}{Sum}}
\\\midrule\textbf{\cellcolor{black}\textcolor{white}{Cost}} &$5.0$&$20.0$&$0.0$&$10.0$&$1.0$&$20.0$&$6.0$&$5.0$&$16.0$&$10.0$&$4.0$&$3.0$&$5.0$&$7.0$&$15.0$&$13.0$&$14.0$&$3.0$&$10.0$&$7.0$&$12.0$&$15.0$&$8.0$&$2.0$&$10.0$&$0.0$&$1.0$&$222.0$
\\\midrule \textbf{\cellcolor{black}\textcolor{white}{Value}} &$10.0$&$20.0$&$4.0$&$17.0$&$3.0$&$20.0$&$15.0$&$9.0$&$20.0$&$16.0$&$20.0$&$10.0$&$6.0$&$8.0$&$8.0$&$10.0$&$6.0$&$10.0$&$20.0$&$20.0$&$15.0$&$20.0$&$20.0$&$5.0$&$0.0$&$0.0$&$0.0$&$312.0$
\\\toprule[1.5pt]
\end{tabular}}
\end{table*}

We simulated value-related dependencies among requirements of a real-world software project (Table~\ref{table_cost_value}) through random generation of qualities and strengths of explicit value-related dependencies. In doing so, for each candidate explicit value-related dependency $(r_i,r_j)$, a random number with uniform distribution in $[-1,1]$ was generated whose sign and magnitude determined the quality and the strength of $(r_i,r_j)$ respectively. A randomly generated $0$ was interpreted as the absence of any explicit value-related dependency from $r_i$ to $r_j$. Experiments were repeated for various levels of value-related dependencies (VDLs) and negative value-related dependencies (NVDLs) among requirements. To achieve this, certain VDLs and NVDLs were imposed on the value-related dependency graph the requirements by randomly removing edges (explicit value-related dependencies) of the graph. Also, experiments were repeated for different percentages of budgets $\%\text{Budget} \in [0,100]$ to investigate efficiency of the experimented release planning models in the presence of various budget constraints. 

\subsection{Simulation Results and Discussion}
\label{sec_simulation_result}

\begin{figure*}[!t]
	\begin{center}
		\subfigure[DA-SRP, NVDL = $0.0$]{%
			\label{fig_sim_ocv_dars_00}
			\includegraphics[scale=0.47]{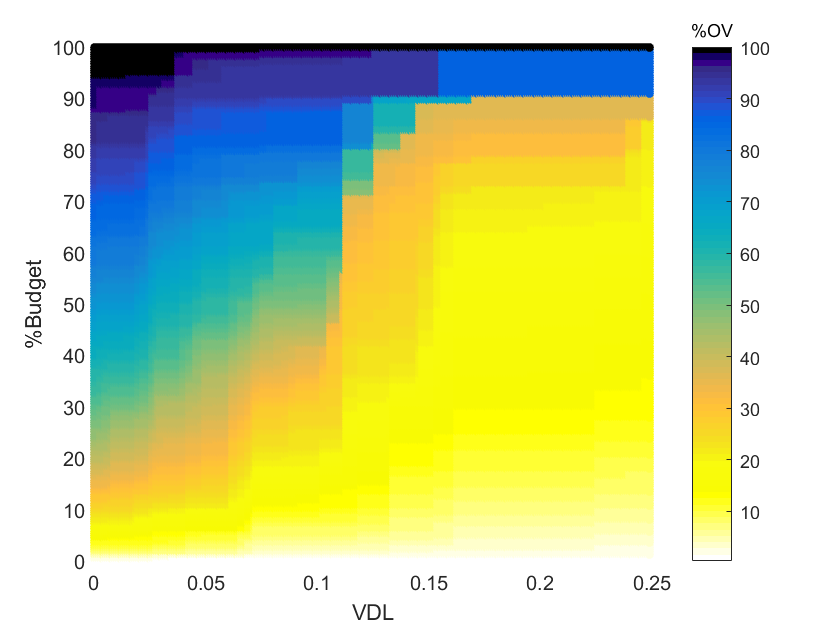}
		}
		\subfigure[BKP, NVDL = $0.0$]{%
			\label{fig_sim_ocv_bkp_00}
			\includegraphics[scale=0.47]{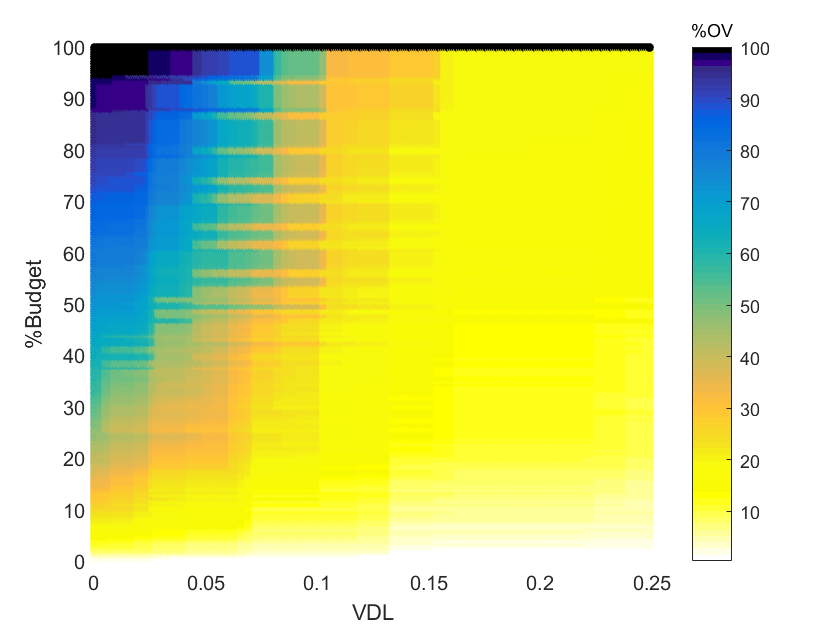}
		}
		\subfigure[BKP-PC, NVDL = $0.0$]{%
			\label{fig_sim_ocv_bkppc_00}
			\includegraphics[scale=0.47]{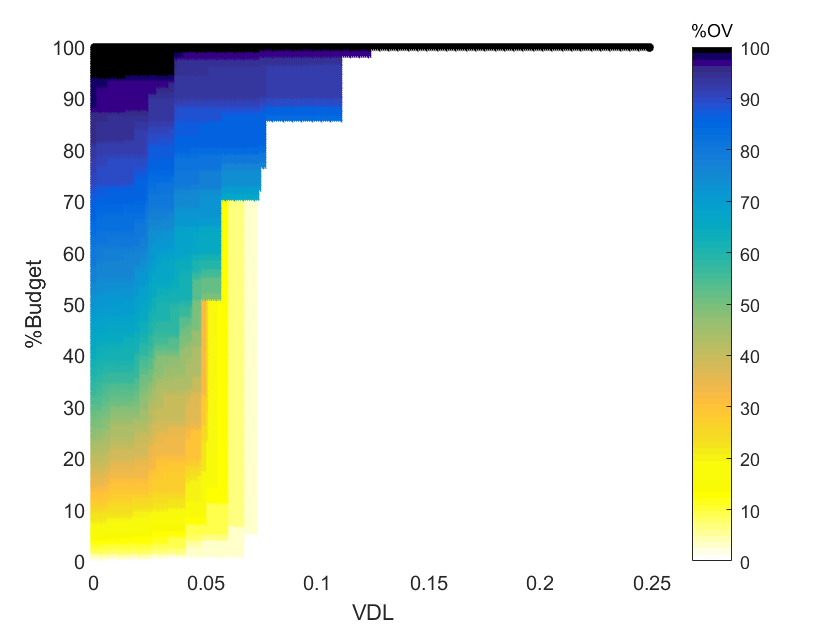}
		}
		\subfigure[DA-SRP, NVDL= $0.5$]{%
			\label{fig_sim_ocv_dars_05}
			\includegraphics[scale=0.47]{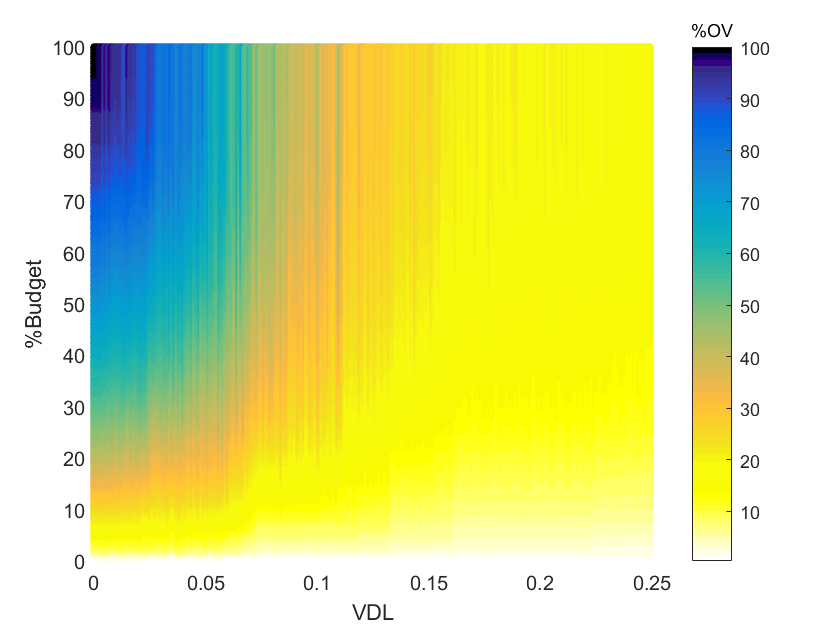}
		}
		\subfigure[BKP, NVDL = $0.5$]{%
			\label{fig_sim_ocv_bkp_05}
			\includegraphics[scale=0.47]{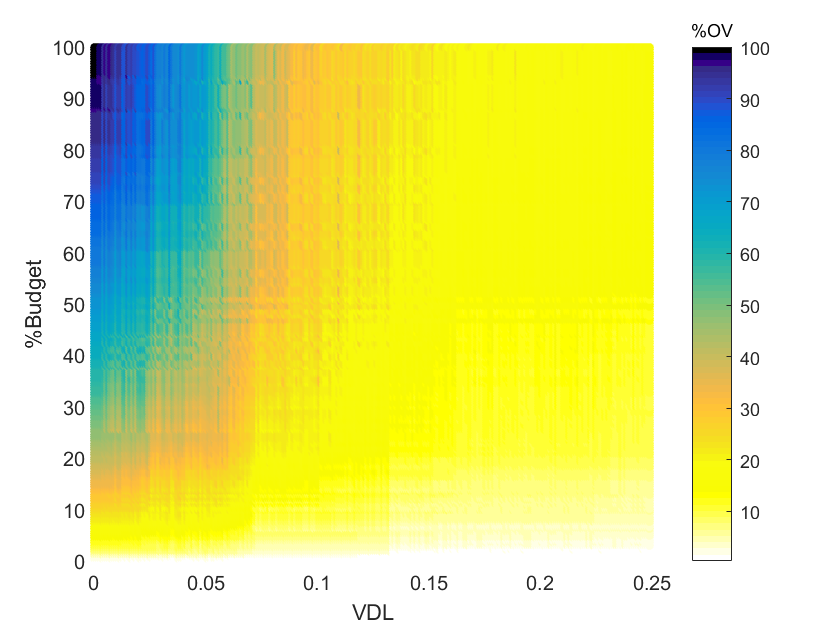}
		}
		\subfigure[BKP-PC, NVDL = $0.5$]{%
			\label{fig_sim_ocv_bkppc_05}
			\includegraphics[scale=0.47]{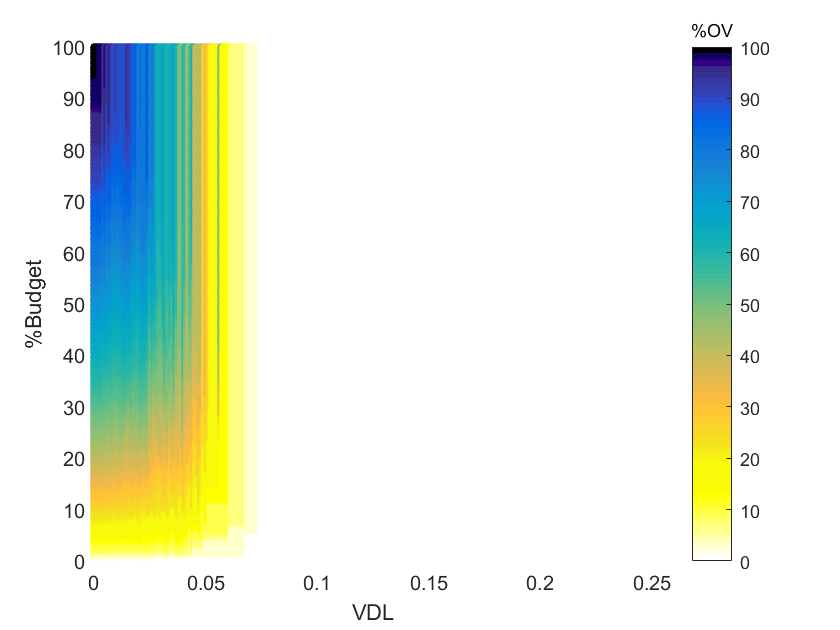}
		}
	\end{center}
	\caption{%
		Overall values achieved from simulated software release planning models.
	}%
	\label{fig_sim_ocv}
\end{figure*}

\begin{figure*}[!t]
	\begin{center}
		\subfigure[DA-SRP, NVDL = $0.0$]{%
			\label{fig_sim_acv_dars_00}
			\includegraphics[scale=0.47]{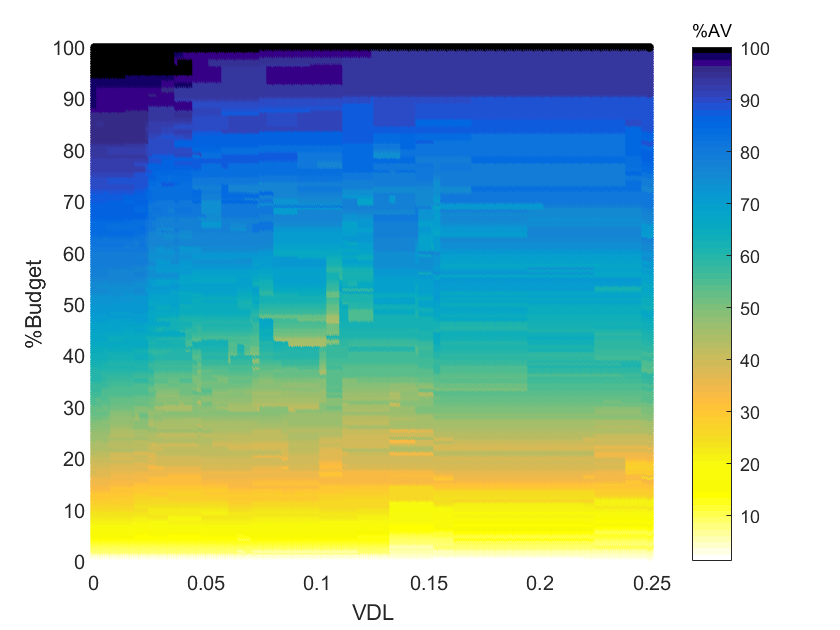}
		}
		\subfigure[BKP, NVDL = $0.0$]{%
			\label{fig_sim_acv_bkp_00}
			\includegraphics[scale=0.47]{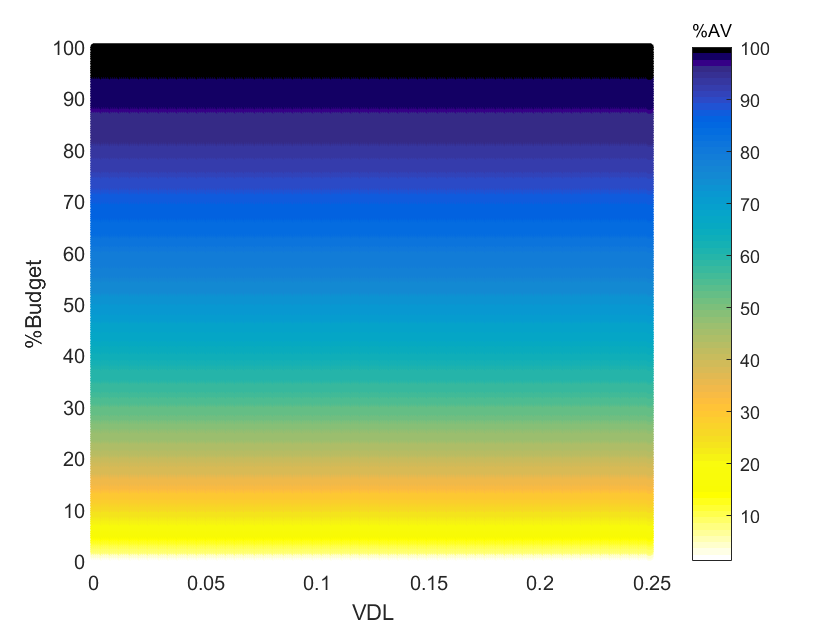}
		}
		\subfigure[BKP-PC, NVDL = $0.0$]{%
			\label{fig_sim_acv_bkppc_00}
			\includegraphics[scale=0.47]{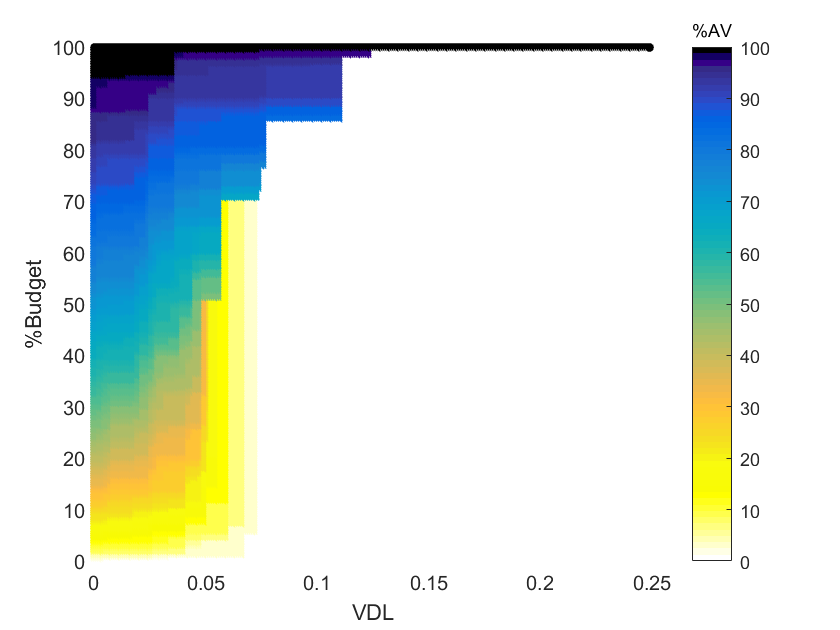}
		}
		\subfigure[DA-SRP, NVDL= $0.5$]{%
			\label{fig_sim_acv_dars_05}
			\includegraphics[scale=0.47]{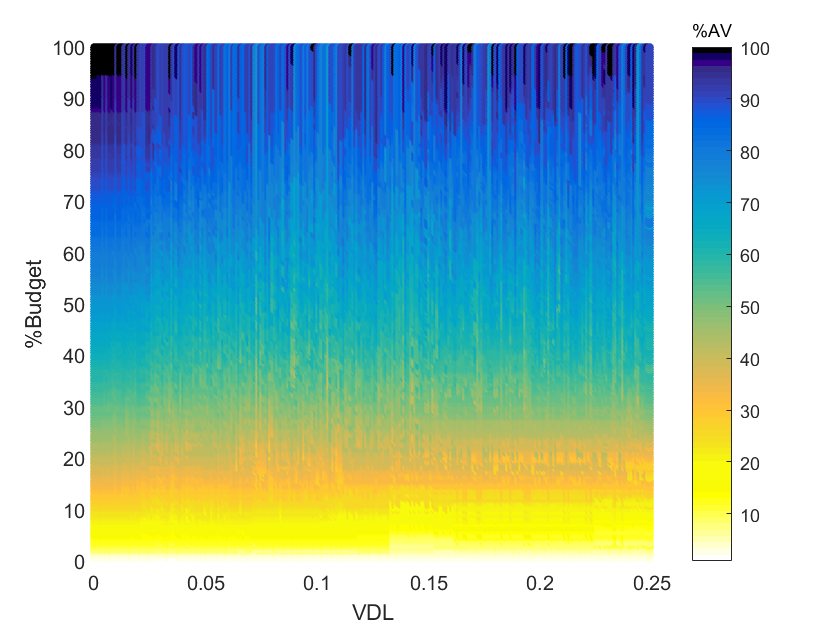}
		}
		\subfigure[BKP, NVDL = $0.5$]{%
			\label{fig_sim_acv_bkp_05}
			\includegraphics[scale=0.47]{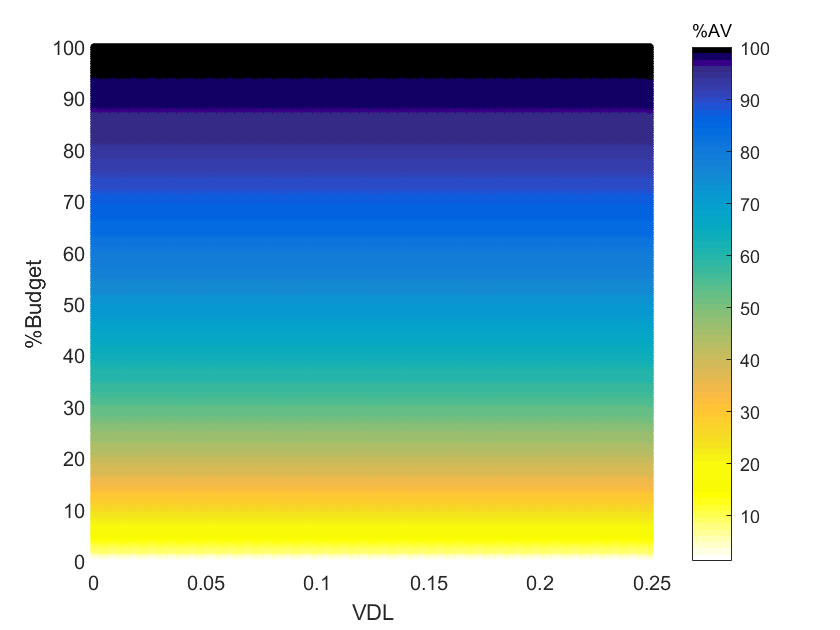}
		}
		\subfigure[BKP-PC, NVDL = $0.5$]{%
			\label{fig_sim_acv_bkppc_05}
			\includegraphics[scale=0.47]{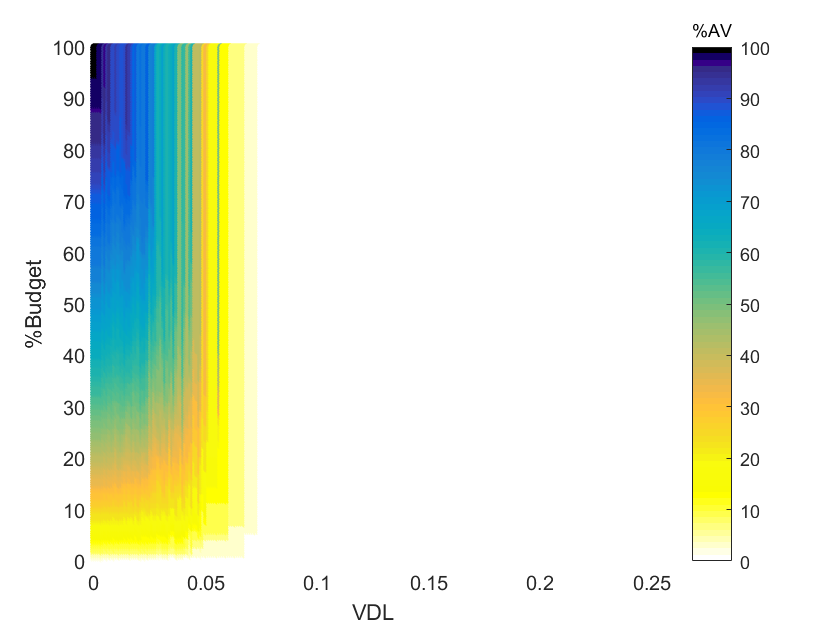}
		}
	\end{center}
	\caption{%
		Accumulated values achieved from simulated software release planning models.
	}%
	\label{fig_sim_acv}
\end{figure*}

Figure~\ref{fig_sim_ocv} and Figure~\ref{fig_sim_acv} show percentages of the overall value ($\%\text{OV}=(\text{OV}/312)\times 100$) and percentages of the accumulated value ($\%\text{AV}=(\text{AV}/312)\times 100$) achieved from the experimented software release planning (SRP) models for two different levels of negative value-related dependencies ($\text{NVDL} = \{0.0,0.5\}$). Give a NVDL, the experiments were repeated for various value dependency levels ($\text{VDL} \in \{0,0.001,...,1\}$), and different percentages of budget ($\%\text{Budget}=(\text{Budget}/222)\times 100$) as denoted by x-axis and y-axis respectively (Figure~\ref{fig_sim_ocv} and Figure~\ref{fig_sim_acv}). For selected subsets of requirements the color-bars in Figure~\ref{fig_sim_ocv} and Figure~\ref{fig_sim_acv} denote the $\%\text{OV}$ and the $\%\text{AV}$ respectively. All our simulation results consistent with the results of our case study showed (Figure~\ref{fig_sim_ocv} and Figure~\ref{fig_sim_acv}) that for all $\%\text{Budget}$, VDLs, and NVDLs, the BKP model always found a subset of requirements with the highest (compared to the BKP-PC and DA-SRP models) accumulated values while overall values of those subsets were not necessarily the highest. The DA-SRP model on the other hand, always found subsets of requirements with the highest overall values while accumulated values of those subsets were not necessarily the highest. 

It was thus, concluded that maximizing the accumulated value of an optimal set and maximizing the overall value of that set are conflicting objectives. Nonetheless, for a given $\%\text{Budget}$, the $\%\text{OV}$ and the $\%\text{AV}$ provided by the BKP and DA-SRP models tended to reduce and converge as NVDL was increased. To formally explain this, more information is needed. However, we can informally state that by increasing the NVDL, the chances that a selected requirement negatively influences the values of other selected requirements, will increase. As such, the chances will increase that ignoring a requirement $r_1$ reduces the value of a requirement $r_2$ (already selected) while selecting $r_1$ reduces the value of a third requirement $r_3$. The impact of increasing NVDL can be clearly observed in Figures~\ref{fig_sim_ocv_dars_05}, \ref{fig_sim_ocv_bkp_05}, \ref{fig_sim_ocv_bkppc_05}, where $\%\text{OV}=100$ has almost never been achieved even in the presence of a sufficient budget. This is more significant in the case of the BKP-PC model (Figure~\ref{fig_sim_ocv_bkppc_05}). The reason is that due to the selection deficiency problem (SDP) in BKP-PC models, increasing negative value-related dependencies will increase the chances that either selecting or ignoring requirements violates the precedence constraints and turn the candidate subset of requirements to an infeasible solution.

We also observed (Figure~\ref{fig_sim_ocv} and Figure~\ref{fig_sim_acv}) in all experiments that overall values given by BKP-PC model were equal to their corresponding accumulated values. The reason is that the BKP-PC model treats all value-related dependencies as binary relations which results in a situation, where a requirement is only selected if all of its positive (negative) value-related dependencies are selected (ignored). As such the penalties of requirements are always calculated as $0$.    

Our results further, showed the selection deficiency problem (SDP)~\cite{mougouei2016factoring} severely impacted the $\%\text{AV}$ and $\%\text{OV}$ achieved from the BKP-PC model in general. This was further exacerbated for higher NVDLs as in such cases not only ignoring but selecting certain requirement resulted in value loss due to the presence of negative value-related dependencies. For instance, for $\text{NVDL}=0.0$ we have $\text{VDL} > 0.12, \%\text{Budget} < 100 \Rightarrow \%\text{AV}(\text{BKP-PC})=\%\text{OV}(\text{BKP-PC})=0$ while for $\text{NVDL}=0.5$ we have $\text{VDL}> 0.07 \Rightarrow \%\text{AV}(\text{BKP-PC})=\%\text{OV}(\text{BKP-PC})=0$. 



Finally, we observed (Figure~\ref{fig_sim_ocv} and Figure~\ref{fig_sim_acv} that) that DA-SRP properly mitigated the impact of SDP by providing some $\%\text{AV}$ and $\%\text{OV}$ when the BKP-PC model was not able to provide any value. The reason is DA-SRP considers the impacts of the ignored (selected) positive (negative) value-related dependencies rather than treating those dependencies as binary  (exists/does not exist) relations. The BKP model however is not subject to the SDP as it completely ignores requirement dependencies all types.

\section{Scalability}
\label{sec_scalability}

Scalability of the work has two main aspects. First, scalability of the dependency identification and second, scalability of our proposed selection model (DA-SRP). Regarding the scalability of the dependency identification, we have automated the process of identifying value-related dependencies based on user preferences for software requirements as described in Section \ref{sec_identification}. Hence, there would be no need for manual comparisons. Our automated dependency identification process is based on the fact that the (expected) value of a requirement is determined by user preferences for that requirement. In this regard, Eells measure of casual strength~\cite{eells1991probabilistic} is used to measure strengths of value-related dependencies based on user preferences. 

The algorithm for computing Eells measure is computationally efficient as it only calculates conditional probabilities for pairs of requirements and their complements. For $n$ requirements hence the complexity is of $O(n^2)$. We have further suggested considering value-implications of precede dependencies ($r_1$ requires/precedes/conflicts-with $r_2$) when such information is available. Precede dependencies are normally identified prior to the release planning as part of requirement analysis. Identifying such dependencies is not specific to our release planning model and therefor does not add any additional overhead to it. 

\begin{figure*}[htbp!]
	\centering
	\centerline{\includegraphics[scale=0.75]{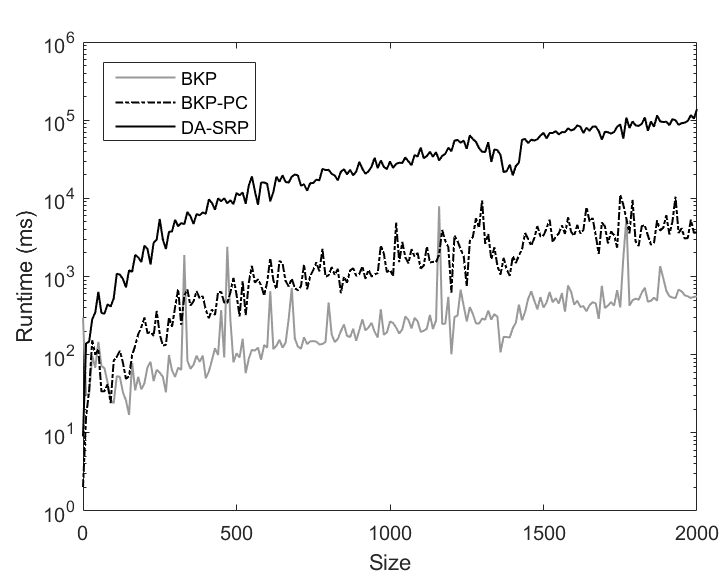}}
	\caption{Runtime of experimented SRP models for different numbers of requirements (size).}
	\label{fig_time}
\end{figure*}

DA-SRP, as given by (\ref{Eq_dars})-(\ref{Eq_dars_c4}), formulates a \textit{Convex Optimization Problem} since the model maximizes a \textit{Concave} objective function over linear constraints~\cite{boyd2004convex}. Convex optimization problems are known to be efficiently solvable~\cite{boyd2004convex}. To further improve this we used the technique introduced in~\cite{mougouei2017bkp} to linearize the DA-SRP model. The model showed to be scalable to large number of requirements as  depicted in simulations results of Figure~\ref{fig_time}). We performed simulations to demonstrate scalability of the DA-SRP model in comparison with the existing \textit BKP and BKP-PC models. Uniformly distributed random numbers were generated to simulate the costs and values of requirements as well as the strengths of value-related dependencies among them. The callable library ILOG CPLEX~\cite{cplex_ibm_2014} was used to implement and run the BKP, BKP-PC, and the DA-SRP models. All simulations were performed on a Windows machine (Windows 8.1) with $8$ GB of RAM and a Core i5-4300 CPU with the speed of 1.9 GHz to 2.49 GHz.

Figure~\ref{fig_time} shows our results with a logarithmic y-axis specifying the run-time of the experimented release planning models and a linear x-axis denoting the size of the experiment (the total number of requirements). Our simulations for random requirement sets of various sizes ($Size = \{0,1,...,2000\}$) demonstrated (Figure~\ref{fig_time}) that the DA-SRP model always found the optimal subset of the requirements (a subset with the highest overall value for a given budget) in less than $2$ minutes. However, the run-time of the DA-SRP model was higher than that of the BKP and BKP-PC models while the BKP model performed faster than the BKP-PC and DA-SRP models (Figure~\ref{fig_time}). The reason is the BKP model ignores value-related dependencies and therefore can be executed in semi-polynomial time using dynamic-programming~\cite{bellman_dynamic_1965} while the BKP-PC needs to satisfy the precedence constraints which capture value-related dependencies and therefore, it takes longer for the BKP-PC model to find the optimal subset of the requirements (solution). The DA-SRP model on the other hand, considers value-related dependencies in addition to the precedence constraints, which increases the execution time of the model and makes it slower than BKP and BKP-PC in general.  

\section{Conclusions and Future Work}
\label{sec_conclusion}

In this paper we presented an integer programming model for software release planning which optimizes the overall value of a software release by considering strengths and qualities of value-related dependencies among software requirements. We further introduced, based on fuzzy graphs, value dependency graphs and their algebraic structure for modeling value-related dependencies and reasoning about qualities and strengths of those dependencies. A modified version of the Floyd-Warshall algorithm was introduced to infer value-related dependencies and their characteristics in polynomial time. Moreover, a scalable technique was proposed for automated identification of explicit value-related dependencies and their characteristics based on user preferences. We also demonstrated application of a latent multivariate Gaussian model for generating samples of large (enough) quantities from user preferences to enhance accuracy of dependency identification. 

We demonstrated through studying an industrial software project and conducting several simulations that: (a) the DA-SRP model can properly capture value-related dependencies in software release planning while being reasonably scalable, (b) the proposed DA-SRP model efficiently mitigates the selection deficiency problem, (c) the DA-SRP model always maximizes the overall value of selected requirements, and (d) maximizing the overall value, with considering value-related dependencies, and the accumulated value, without considering value-related dependencies, of a selected subset of software requirements are conflicting objectives.

This work can be extended in several directions. One is to explore different measures of casual strength and comparing their efficiency in capturing value-related dependencies. To assist dependency identification, one may also consider mining online resources such as social media and software repositories. In this paper we studied value as the economic worth of a software. But, it is also important to consider social aspects of software and their impacts on the society. Considering social values of software are particularly important to stakeholders with special preferences e.g. users with disabilities. It would be interesting hence to integrate social values in software through considering those values in software release planning. In this regard, considering dependencies/conflicts among social values and reconciling those into software would be imperative. 

\bibliography{ref}
\end{document}